\documentclass{aa}

\usepackage{txfonts}
\usepackage{longtable}
\usepackage{rotating}
\usepackage{natbib}
\usepackage{graphicx}
\usepackage{graphics}
\usepackage{psfrag}
\usepackage{amssymb}
\usepackage{xspace}
\usepackage{color}
\bibpunct{(}{)}{;}{a}{}{,}
\def\Teff{$T_{\mathrm{eff}}$}

\def\kms{$\mathrm{km\,s}^{-1}$}

\def\Ro{\ensuremath{R_{\odot}}}
\def\Rs{\ensuremath{R_{\rm star}}}

\def\Mo{\ensuremath{M_{\odot}}}
\def\Ms{\ensuremath{M_{\rm star}}}

\def\Re{\ensuremath{R_{\oplus}}}
\def\Me{\ensuremath{M_{\oplus}}}
\def\Rt{\ensuremath{R_{\rm T}}}
\def\Rpl{\ensuremath{R_{\rm pl}}}
\def\Rrl{\ensuremath{R_{\rm RL}}}
\def\Mpl{\ensuremath{M_{\rm pl}}}
\def\tb{$\tilde\lambda$}
\def\tbs{$\tilde\lambda^*$}
\def\b{$\lambda$}
\def\bs{$\lambda^*$}
\def\B{$\Lambda$}
\def\Bc{$\Lambda_{\rm T}$}

\def\ergscm{erg\,s$^{-1}$\,cm$^{-2}$}

\def\teq{$T_{\rm eq}$}
\def\len{$L_{\rm en}$}
\def\lhy{$L_{\rm hy}$}
\def\Reff{$R_{\rm XUV_{eff}}$}

\begin{document}
\title{Aeronomical constraints to the minimum mass and maximum radius of hot low-mass planets}
\subtitle{}
\author{L. Fossati\inst{1} 			\and
	N. V. Erkaev\inst{2}			\and
	H. Lammer\inst{1}			\and
	P. E. Cubillos\inst{1}			\and
	P. Odert\inst{1}			\and
	I. Juvan\inst{1}			\and
	K. G. Kislyakova\inst{1}		\and
	M. Lendl\inst{1,3}			\and
	D. Kubyshkina\inst{1}			\and
	S. J. Bauer\inst{4}
}
\institute{
	Space Research Institute, Austrian Academy of Sciences, Schmiedlstrasse 		6, A-8042 Graz, Austria\\
	\email{luca.fossati@oeaw.ac.at}
	\and
	Federal Research Center ``Krasnoyarsk Science Center'' SB RAS, 				``Institute of Computational Modelling'', Krasnoyarsk 36, Russian 			Federation
	\and
	Max Planck Institute for Astronomy, K\"onigstuhl 17, 69117 -- 				Heidelberg, Germany	
	\and
	Institut f\"ur Geophysik, Astrophysik und Meteorologie, 				Karl-Franzens-Universit\"at, Universit\"atsplatz 5, 8010 Graz, Austria
}
\date{}
\abstract
{Stimulated by the discovery of a number of close-in low-density planets, we generalise the Jeans escape parameter taking hydrodynamic and Roche lobe effects into account. We furthermore define $\Lambda$ as the value of the Jeans escape parameter calculated at the observed planetary radius and mass for the planet's equilibrium temperature and considering atomic hydrogen, independently of the atmospheric temperature profile. We consider 5 and 10\,\Me\ planets with an equilibrium temperature of 500 and 1000\,K, orbiting early G-, K-, and M-type stars. Assuming a clear atmosphere and by comparing escape rates obtained from the energy-limited formula, which only accounts for the heating induced by the absorption of the high-energy stellar radiation, and from a hydrodynamic atmosphere code, which also accounts for the bolometric heating, we find that planets whose $\Lambda$ is smaller than 15--35 lie in the ``boil-off'' regime, where the escape is driven by the atmospheric thermal energy and low planetary gravity. We find that the atmosphere of hot (i.e. \teq\,$\gtrapprox$\,1000\,K) low-mass (\Mpl\,$\lessapprox$\,5\,\Me) planets with $\Lambda$\,$<$\,15--35 shrinks to smaller radii so that their $\Lambda$ evolves to values higher than 15--35, hence out of the boil-off regime, in less than $\approx$500\,Myr. Because of their small Roche lobe radius, we find the same result also for hot (i.e. \teq\,$\gtrapprox$\,1000\,K) higher mass (\Mpl\,$\lessapprox$\,10\,\Me) planets with $\Lambda$\,$<$\,15--35, when they orbit M-dwarfs. For old, hydrogen-dominated planets in this range of parameters, $\Lambda$ should therefore be $\geq$15--35, which provides a strong constraint on the planetary minimum mass and maximum radius and can be used to predict the presence of aerosols and/or constrain planetary masses, for example.}
\titlerunning{Constrain the parameters of hot low-mass planets}
\authorrunning{L. Fossati et al.}
\maketitle
\section{Introduction}\label{sec:intro}
Thanks to the large number of extra-solar planets (exoplanets) discovered to date by ground- and space-based facilities, such as SuperWASP \citep{pollacco2006}, HATNet \citep{bakos2004}, CoRoT \citep{corot}, Kepler \citep{kepler}, and K2 \citep{howell2014}, we are beginning to classify the large variety of detected exoplanets on the basis of their properties. One of the greatest recent surprises in planetary sciences was the discovery of a large population of planets with mass and radius in between that of terrestrial and giant planets of the solar system \citep{mullally2015}. These planets, hereafter sub-Neptunes, typically have masses and radii in the 1.5--17\,\Me\ and 1.5--5\,\Re\ range. Sub-Neptunes fill a gap of physical parameters that are absent from the solar system. Accurately deriving their masses and radii is therefore crucial to our overall understanding of planets.

The high quality of the Kepler light curves allowed us to obtain precise transit radii, even for small planets, but for most of them, the low mass and faint apparent magnitude of their host stars hampers a precise enough determination of the planetary mass through radial velocity. For several multi-planet systems, planetary masses have been inferred from transit-timing variations (TTVs), but some of the resulting values are at odds with those derived from radial velocity \citep[e.g.,][]{weiss2014}. Sub-Neptunes for which both mass and radius have been measured present a large spread in bulk density ($\approx$0.03--80\,g\,cm$^{-3}$; low average densities imply the presence of hydrogen-dominated atmospheres), which finding is currently greatly debated \citep[e.g.,][]{lopez2012,howe2014,howe2015,lee2015,lee2016,owen2015,ginzburg2015}.

It is therefore important to find external independent constraints to planetary masses and radii that could be applied to a large number of planets, for example to independently test the masses derived from TTVs, identify the possible presence of high-altitude aerosols, and estimate a realistic range of planetary radii/masses given a certain mass/radius. We show here how basic aeronomical considerations, supported by hydrodynamic modelling and previous results \citep{owen2016}, can constrain the mass/radius of old sub-Neptunes given their radius/mass and equilibrium temperature (\teq).
\section{Generalisation of the Jeans escape parameter}\label{sec:beta}
The Jeans escape parameter \b\ is classically defined at the exobase and for a hydrostatic atmosphere. It is the ratio between the escape velocity $\upsilon_{\infty}$ and the most probable velocity $\upsilon_{\rm 0}$ of a Maxwellian distribution at temperature $T$, squared \citep{jeans1925,chamber1963,opik1963,bauer2004}. We generalise the Jeans escape parameter at each atmospheric layer $r$ and corresponding temperature $T$ for a hydrodynamic atmosphere composed of atomic and molecular hydrogen as
\begin{equation}
\begin{split}
\lambda^*(r) \equiv \frac{\upsilon_{\infty}^2}{\upsilon_{\rm 0}^2} = \frac{\upsilon_{\infty}^2}{\left(\upsilon_{\rm hy}/2+\sqrt{\upsilon_{\rm hy}^2/4+\upsilon_{\rm th}^2}\right)^2} = \\ \frac{2GM_{\rm pl}}{r\left(\upsilon_{\rm hy}/2+\sqrt{\upsilon_{\rm hy}^2/4+2k_{\rm B}T/m}\right)^2}\,,
\label{eq:beta}
\end{split}
\end{equation}
where $G$ is Newton's gravitational constant, $k_{\rm B}$ is Boltzmann's constant, \Mpl\ is the planetary mass, $\upsilon_{\rm th}$ is the thermal velocity $\sqrt{2k_{\rm B}T/m}$, and $\upsilon_{\rm hy}$ is the bulk velocity of the particles at each atmospheric layer. In Eq.~\ref{eq:beta}, $m$ is the mean molecular weight
\begin{equation}
m = \frac{\sum n_{\rm X}m_{\rm X}}{\sum n_{\rm X}}
\label{eq:m}
\end{equation}
where $n_{\rm X}$ and $m_{\rm X}$ are the density and mass of each atom/molecule (X) in the atmosphere. In this work, we consider atomic and molecular hydrogen.

The value of $\upsilon_{\rm 0}$ in the hydrodynamic case is that of a shifted Maxwellian distribution, where $\upsilon_{\rm hy}$ is the shift. The Maxwellian velocity distribution gives the number of particles between $\upsilon$ and $\upsilon+d\upsilon$ and can be written as
\begin{equation}
\label{eq:maxwell}
F(\upsilon)d\upsilon = 4 \pi n \left(\frac{m}{2 \pi k_{\rm B}T}\right)^{3/2}\upsilon^2 \exp\left(-\frac{m(\upsilon-\upsilon_{\rm hy})^2}{2k_{\rm B}T}\right) d\upsilon,
\end{equation}
where $n$ is the number density and $m$ the particle mass. The most probable velocity $\upsilon_0$ is found where Eq.~(\ref{eq:maxwell}) has its maximum and can therefore be derived by setting $dF/d\upsilon$\,=\,0. This condition results in a quadratic equation for $\upsilon$,
\begin{equation}
\frac{\upsilon^2}{\upsilon_{\rm th}^2} - \frac{\upsilon\,\upsilon_{\rm hy}}{\upsilon_{\rm th}^2} - 1 = 0.
\end{equation}
The solution of this equation is
\begin{equation}
\label{eq:v0}
\upsilon_0 = \frac{\upsilon_{\rm hy}}{2} + \sqrt{\frac{\upsilon_{\rm hy}^2}{4} + \upsilon_{\rm th}^2},
\end{equation}
where only this positive solution is physical (the negative solution yields a negative $\upsilon_0$). Note that a direct derivation of $\upsilon_0$ by setting $\upsilon_{\rm hy}=0$ in Eq.~(\ref{eq:maxwell}) or in Eq.~(\ref{eq:v0}) yields $\upsilon_0 = \upsilon_{\rm th}$. From Eq.~(\ref{eq:v0}) it also follows that if $\upsilon_{\rm th}\rightarrow 0$, then $\upsilon_0\rightarrow \upsilon_{\rm hy}$, as expected.

The formulation of the Jeans escape parameter given in Eq.~\ref{eq:beta} is reminiscent of the ``solar breeze'' used before Parker's solar wind model was accepted \citep[e.g.,][]{chamberlain1960,chamberlain1961}. If $\upsilon_{\rm hy}$ is negligible compared to $\upsilon_{\rm th}$ (i.e. hydrostatic atmosphere), the Jeans escape parameter returns to the classical form of
\begin{equation}
\lambda^* = \lambda = \frac{GM_{\rm pl}m}{k_{\rm B}Tr}\,.
\end{equation}
We recall that for the classical Jeans escape parameter (hydrostatic atmosphere), a layer is completely bound to a planet for \b\,$\gtrsim$\,30 and escape is important for \b\,$<$\,15, while for \b\,$\lesssim$\,1.5 the atmosphere is in hydrodynamic ``blow-off'' \citep{jeans1925,chamber1963,opik1963,bauer2004}. This last condition occurs when the thermal energy of the gas is very close to, or even exceeds, the gravitational energy.

The vast majority of the exoplanets known to date orbits at close distance to their host stars. We therefore consider Roche-lobe effects. Following the procedure described in Sect.~2 of \citet{erkaev2007}, in Eq.~(\ref{eq:beta}) we substitute the gravitational potential difference between the planetocentric distance $r$ and infinity ($GM_{\rm pl}/r$) by the gravitational potential difference between $r$ and the Roche-lobe radius ($\Delta\phi$). We therefore obtain
\begin{equation}
\tilde{\lambda}^*(r) = \frac{2\Delta\phi}{\left(\upsilon_{\rm hy}/2+\sqrt{\upsilon_{\rm hy}^2/4+2k_{\rm B}T/m}\right)^2}\,,
\label{eq:betanew}
\end{equation}
where
\begin{equation}
\Delta\phi = \phi_{\rm 0}\frac{\xi-1}{\xi}\left[1-\frac{1}{\delta}\frac{\xi}{\gamma^2}\frac{\gamma(1+\xi)-\xi}{(\gamma-1)(\gamma-\xi)}-\frac{\xi(1+\delta)(1+\xi)}{2\delta\gamma^3}\right]
\label{eq:deltaphi}
\end{equation}
\citep[see Eq.~(7) of][]{erkaev2007} and
\begin{equation}
\phi_{\rm 0} = G\frac{M_{\rm pl}}{r}\,,\,\,\,\delta = \frac{M_{\rm pl}}{M_{\star}}\,,\,\,\,\gamma = \frac{d}{r}\,,\,{\rm and}\,\,\,\xi = \frac{R_{\rm RL}}{r}\,.
\label{eq:varia}
\end{equation}
In Eq.~(\ref{eq:varia}), $M_{\star}$ is the stellar mass, $d$ is the semi-major axis, and \Rrl\ is the Roche lobe radius. Therefore, Eq.~\ref{eq:betanew} gives the generalised form of the Jeans escape parameter.
\subsection{Planet atmosphere modelling}\label{sec:modelling}
%
\begin{figure*}
\centerline{\includegraphics[width=18.0cm,clip]{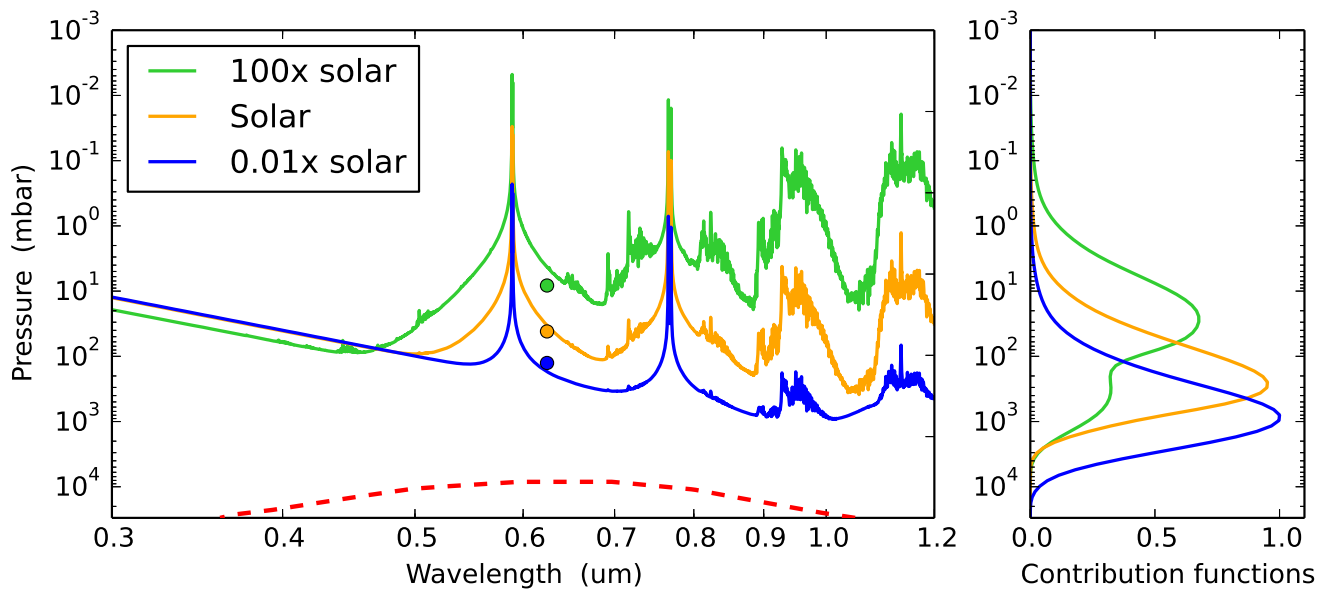}}
\caption{Left: synthetic transmission spectra calculated for a planet with \Mpl\,=\,5\,\Me, \Rpl\,=\,4\,\Re, and a 1000\,K isothermal atmosphere with 0.01 (blue), 1.0 (orange), and 100 (green) times solar metallicity. The circles of corresponding colour denote the transmission curves integrated over the CoRoT spectral response curve (red dashed curve). Right: contribution functions for the vertical optical depth integrated over the CoRoT spectral response curve.}
\label{fig:spectra}
\end{figure*}
To draw profiles of \bs\ and \tbs\ we derive the temperature, pressure, velocity, and density structure of planetary atmospheres employing a stellar high-energy (XUV; 1--920\,\AA) absorption and 1D hydrodynamic upper-atmosphere model that solves the system of hydrodynamic equations for mass, momentum, and energy conservation, and also accounts for ionisation, dissociation, recombination, and Ly$\alpha$ cooling. The full description of the hydrodynamic code adopted for the simulations is presented in \citet{erkaev2016}.

Hydrodynamic modelling is valid in presence of enough collisions, which occurs for Knudsen number $Kn=l/H<0.1$ \citep{volkov2011}, where $l$ is the mean free path and $H$ is the local scale height; in the domain of our models, from \Rpl\ to \Rrl, this criterion is always fulfilled. Throughout our calculations, we adopt a net heating efficiency ($\eta$) of 15\% \citep{shematovich2014} and use stellar XUV fluxes ($I_{\rm XUV}$) estimated from the average solar XUV flux \citep{ribas2005}, scaled to the appropriate distance and stellar radius. We note that X-ray heating is not relevant in our case, because we do not consider active young stars \citep{owen2012}. We also assumed that at \Rpl\ hydrogen is completely in molecular form (i.e. H$_2$), which is true for planets with \teq\,$<$\,2000\,K \citep{koskinen2010}.

For all calculations, and throughout the paper, we consider that \Rpl\ lies at a fiducial atmospheric pressure ($p_0$) of 100\,mbar. To justify this assumption, we calculated the photospheric deposition level using an updated version of the radiative transfer code described in \citet{Cubillos2016phd} and \citet{Blecic2016pdh}. The model considers opacities from line-by-line transitions from HITEMP for H$_2$O, CO, and CO$_2$ \citep{hitemp} and HITRAN for CH$_4$ \citep{hitran}. In addition, it includes opacities for H$_2$--H$_2$ and H$_2$--He collision-induced absorption from \citet{borisov2002}, \citet{borisov2001}, and \citet{Jorgensen2000}, H$_2$ Rayleigh scattering from \citet{lecav08}, and sodium and potassium doublets from \citet{burrows2000}.

\begin{figure}
\centerline{\includegraphics[width=9.0cm,clip]{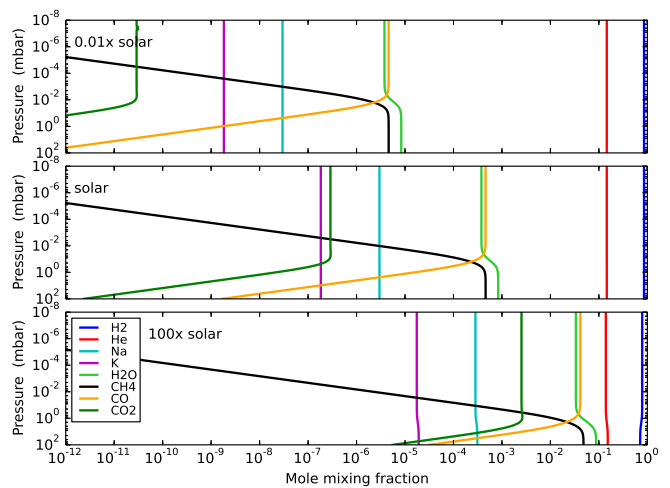}}
\caption{Mole-mixing fractions of the atmospheric species (see legend in the bottom panel) in thermochemical equilibrium for isothermal (1000\,K) models calculated for 0.01 (top), 1.0 (middle), and 100 (bottom) times solar metallicity.}
\label{fig:atmosphere}
\end{figure}

In Fig.~\ref{fig:spectra} we present transmission spectra for a fiducial sub-Neptune with \Mpl\,=\,5\,\Me, \Rpl\,=\,4\,\Re, and an isothermal atmosphere at 1000\,K, in hydrostatic and thermochemical equilibrium. We explored three different cases varying the atmospheric elemental metallicities, considering 0.01, 1.0, and 100 times solar abundances (Figs.~\ref{fig:spectra} and \ref{fig:atmosphere}). We adjusted the pressure-radius reference level such that the resulting transmission radius (integrated over the optical band) matches the fiducial planetary radius, adopting the CoRoT spectral response curve, as an example. We find that the planetary transmission radii correspond to pressure levels of 130, 50, and 10\,mbar for the 0.01, 1.0, and 100.0$\times$ solar-metallicity models, respectively (Fig.~\ref{fig:spectra}, left panel).

After we obtained the pressure-radius relationship, we computed the contribution functions (in the optical band) for the vertical optical depth. The barycenter (i.e., average) of the contribution functions indicate where the atmosphere becomes optically thick. This is the position of the planetary photosphere, where the lower boundary for the hydrodynamic calculation would need to be set. We find that for the planet considered here the photospheric deposition level is approximately located at 551, 159, and 33\,mbar for the 0.01, 1.0, and 100.0$\times$ solar-metallicity models, respectively (Fig.~\ref{fig:spectra}, right panel).

We performed the same procedure for all planets analysed in this work and list the pressure corresponding to the barycenter of the contribution function in the fifth column of Table~\ref{tab:runs}. The pressure values range between about 100 and 700 mbar, where the lower pressure values are obtained for the cooler, lower density planets. Figure~\ref{fig:spectra} shows that a higher metallicity, as expected for low-mass planets, would lead to a slight decrease in pressure values, hence justifying our assumption of placing \Rpl\ at an average 100\,mbar pressure level.

Our hydrodynamic model implicitly considers the stellar continuum absorption by setting the temperature at the lower boundary, hence at \Rpl\  (i.e. where most of the stellar radiation is absorbed), equal to \teq. We return to the validity of this approximation in Sect.~\ref{sec:betacritical}. The planets considered here are old, hence heating from the planet interior can be neglected. 
\subsection{\bs\ and \tbs\ profiles}\label{sec:profiles}
\begin{figure}
\centerline{\includegraphics[width=9.0cm,clip]{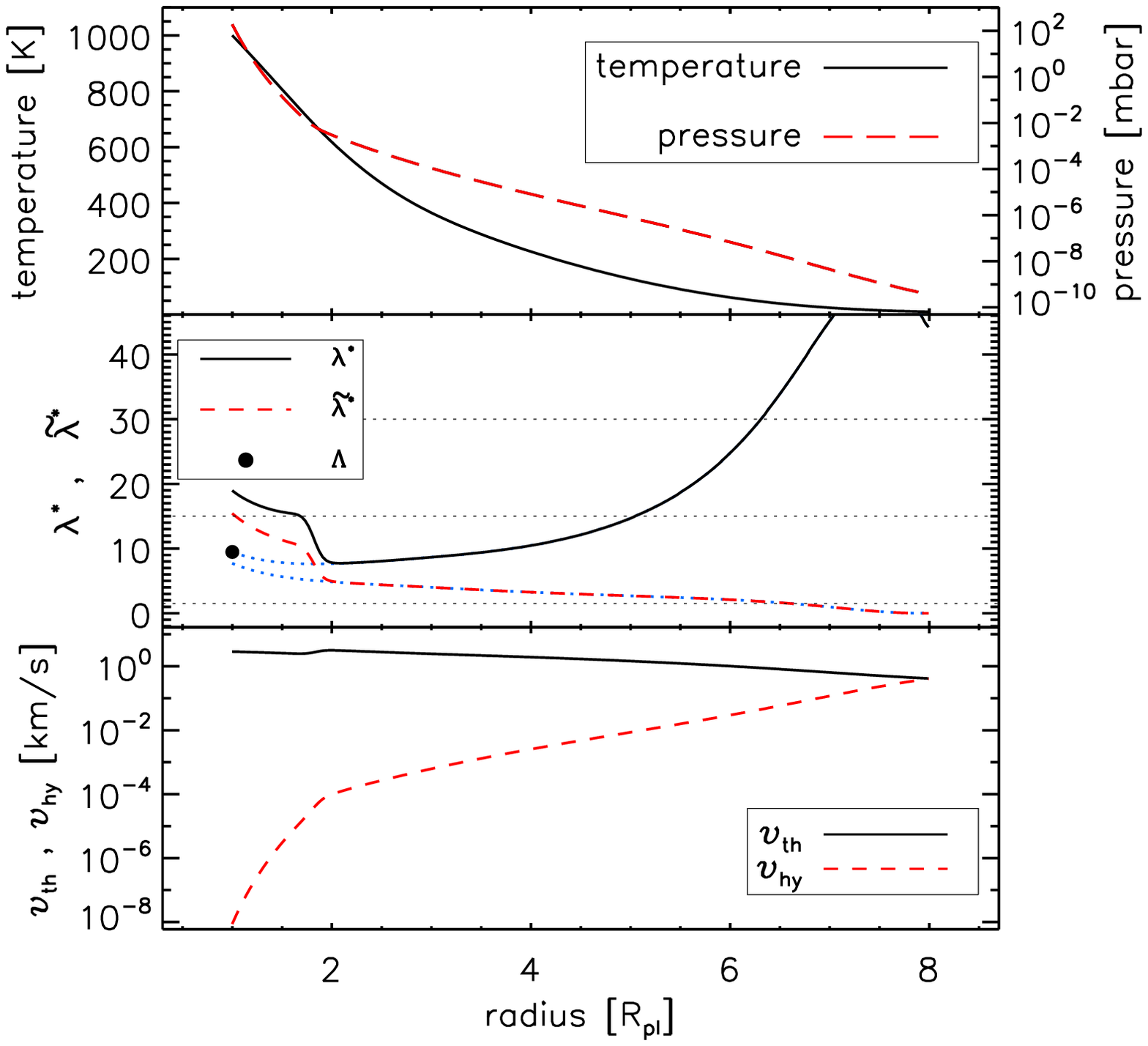}}
\caption{Top: temperature (black solid line) and pressure (red dashed line) profiles as a function of radius $r$ in units of \Rpl\ for a 5\,\Me\ and 4\,\Re\ planet with \teq\,$=1000$\,K, orbiting an early K-type star (see Table~\ref{tab:runs}). The right axis indicates the pressure scale. Middle: \bs\ (black solid line) and \tbs\ (red dashed line) profiles as a function of radius $r$ in units of \Rpl. The horizontal lines mark the critical values of the Jeans escape parameter in the hydrostatic case: 1.5, 15, and 30. The blue dotted lines show the \bs\ and \tbs\ profiles calculated assuming that the whole atmosphere is made of atomic hydrogen. The filled circle indicates the \B\ value (see Sect.~\ref{sec:betacritical}). Bottom: $\upsilon_{\rm th}$ (black solid line) and $\upsilon_{\rm hy}$ (red dashed line) profiles in \kms.}
\label{fig:lowbeta}
\end{figure}

As an example to show the differences between \bs\ and \tbs, we modelled a close-in low-density 5\,\Me\ and 4\,\Re\  (average density $\rho$ of 0.4\,g\,cm$^{-3}$) planet with \teq\ of 1000\,K, orbiting an early K-type star (see Table~\ref{tab:runs}). The parameters adopted for this idealised planet are similar to those of Kepler-87c \citep{ofir2014}. We derived the mean molecular mass at each atmospheric layer from the modelled H and H$_2$ mixing ratios. Figure~\ref{fig:lowbeta} shows the obtained profiles.

In the 1--2\,\Rpl\ range, \bs\ decreases with increasing $r$ because the gravitational potential decreases and the H$_2$ molecules dissociate under the action of the stellar XUV flux. All H$_2$ molecules are dissociated at $\sim$2\,\Rpl. Then, at larger radii, as the temperature continues to decrease due to adiabatic cooling, \bs\ increases and remains above 30 for radii grater than 6.5\,\Rpl. This implies that no particles could escape, regardless of their proximity to \Rrl, which is non-physical. Instead, \tbs\ monotonically decreases with increasing $r$. The bottom panel of Fig.~\ref{fig:lowbeta} shows that for such a close-in planet, despite the hydrodynamic nature of the atmosphere, in most layers $\upsilon_{\rm hy}$ is negligible compared to $\upsilon_{\rm th}$, therefore \bs\,$\approx$\,\b\ and \tbs\,$\approx$\,\tb.

Figure~\ref{fig:lowbeta} shows that the value of \tbs\ approaches unity at atmospheric layers where the pressure, hence density, is high enough to power high escape rates (see Table~\ref{tab:runs}). These upper layers are in a blow-off regime where the escaping gas is continuously replenished by the hydrodynamically expanding atmosphere, with the expansion being driven by the high thermal energy and low planet gravity. This escape regime, here presented from an aeronomical point of view, has been discovered and thoroughly described by \citet{owen2016}, who called it ``boil-off'', in relation to the study of the evolution of young planets that are just released from the protoplanetary nebula \citep[see also][]{ginzburg2015}.
\section{Using escape rates to identify planets in the boil-off regime}\label{sec:betacritical}
We define \B\ as the Jeans escape parameter \b\  (without accounting for Roche-lobe effects and hydrodynamic velocities) at \Rpl, evaluated at the \teq\ of the planet and for an atomic-hydrogen gas (see the full dot and the blue dotted lines in Fig.~\ref{fig:lowbeta})
\begin{equation}
\Lambda = \frac{GM_{\rm pl}m_{\rm H}}{k_{\rm B}T_{\rm eq}R_{\rm pl}}\,. 
\end{equation}
This quantity, which we call the restricted Jeans escape parameter, is useful because it can be derived for any planet for which mass, transit radius, and \teq\ are measured, and without the need of any atmospheric modelling or calculation of \Rrl. We aim here at roughly finding the threshold \B\ values (\Bc), as a function of \Mpl, \Rpl, and \teq, below which the atmosphere transitions towards the boil-off regime. For this we use escape rates, as described below.

In addition to the escape rates derived from the hydrodynamic model (\lhy), we consider the maximum possible XUV-driven escape rates, which can be analytically estimated using the energy-limited formula \citep[e.g.,][]{watson1981,erkaev2007},
\begin{equation}
L_{\rm en}=\frac{\pi\eta R_{\rm pl}R_{\rm XUV_{eff}}^2I_{\rm XUV}}{GM_{\rm pl}m_{\rm H}K(\xi)}\,,
\label{eq:energyLimited}
\end{equation}
where \Reff\ is the effective radius at which the XUV energy is absorbed in the upper atmosphere \citep[see Table~\ref{tab:runs};][]{erkaev2007,erkaev2015} and $\eta$ is the heating efficiency (see Sect.~\ref{sec:modelling}). The factor $K(\xi)=1-\frac{3}{2\xi}+\frac{1}{2\xi^3}$ accounts for Roche-lobe effects \citep{erkaev2007}. We note that Roche-lobe effects are also considered in the hydrodynamic model.

By construction, XUV heating and the intrinsic thermal energy of the atmosphere are considered in the computation of \lhy, while only XUV heating is taken into account when deriving \len. It follows that the boil-off regime, that is, when the intrinsic thermal energy of the atmosphere becomes the efficient main driver of the escape, occurs for \lhy\ greater than \len. For this situation, \lhy/\len\,$>1$ cannot be achieved purely from XUV heating, implying that the outflow must be driven by the heat present at the lower boundary of the atmosphere. We can therefore use the \lhy/\len\,$\approx1$ as an empirical condition to estimate \Bc.

\begin{figure}
\centerline{\includegraphics[width=9.0cm,clip]{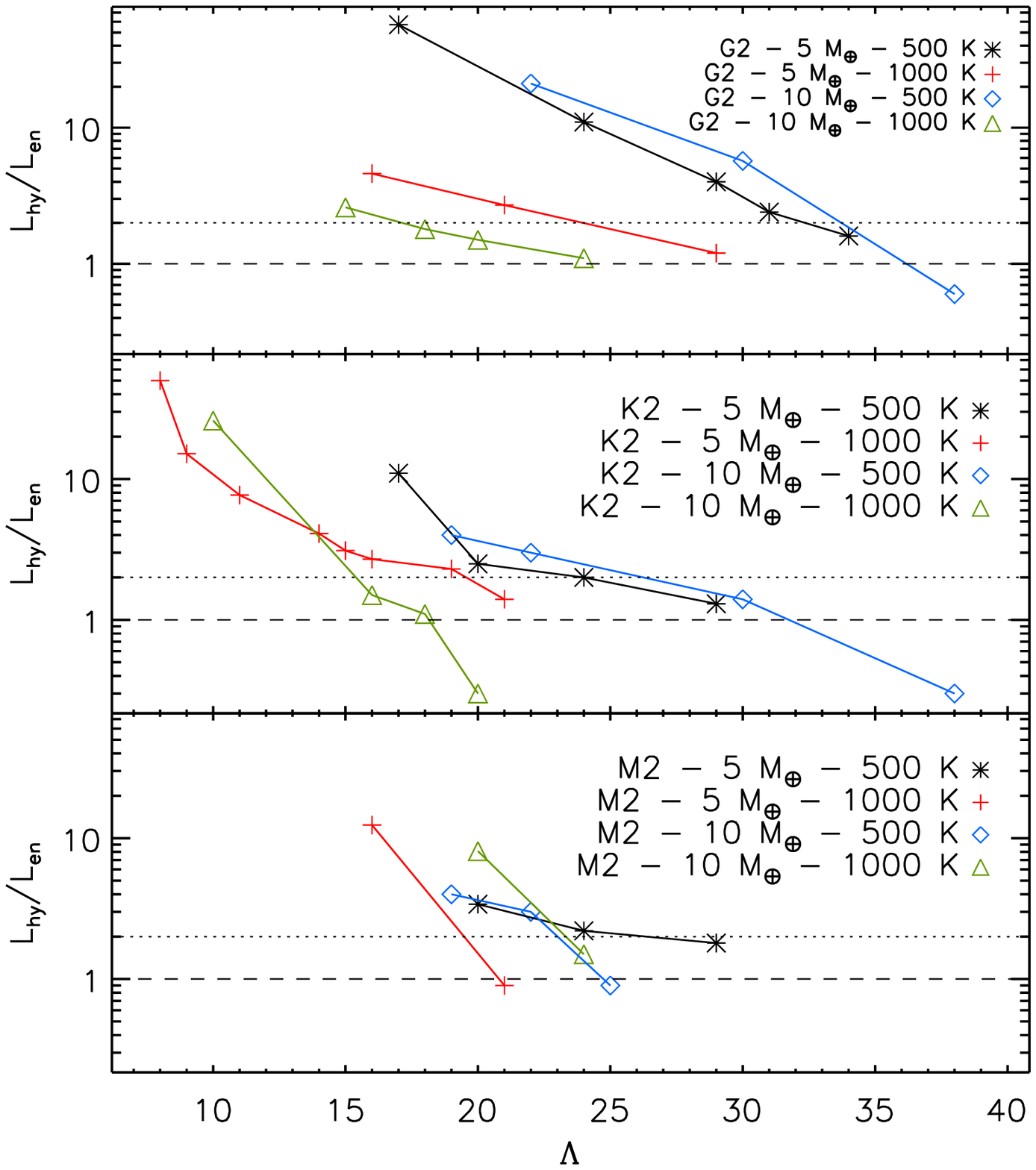}}
\caption{Ratio between the hydrodynamic (\lhy) and energy-limited (\len) escape rates as a function of \B\ for the modelled planets orbiting the G2 (top), K2 (middle), and M2 (bottom) star. Within each panel, the legend indicates the mass (in \Me) and temperature (in K) of the modelled planets. The dashed line indicates the equality between \lhy\ and \len, while the dotted line indicates where the \lhy/\len\ ratio is equal to 2.0. The value of \Bc\ lies between 15 and 35.}
\label{fig:table-plot}
\end{figure}

To identify the \Bc\ value, which is the \B\ value satisfying the \lhy/\len\,$\approx1$ condition, we ran a set of hydrodynamic simulations for two idealised old planets of 5 and 10\,\Me\ orbiting an early G-, K-, and M-type star at distances such that \teq\ is equal to 500 and 1000\,K, assuming a Bond albedo of 0.3. Table~\ref{tab:runs} lists the complete set of input parameters and results, which are visually displayed in Fig.~\ref{fig:table-plot}.
\begin{table*}[ht!]
\caption{Input parameters and results of the simulations performed with $\eta$\,=\,15\%.}
\label{tab:runs}
\begin{center}
\begin{tabular}{ccccccccccccr}
\hline
\Mpl & \Rpl & $\rho$       & \teq & $p_0$ & $I_{\rm XUV}$ & $d$ & \B & Roche lobe       & \Reff & \lhy     & \len     & \lhy/\len \\
\Me  & \Re  & g\,cm$^{-3}$ & K    &  mbar   & \ergscm       & AU  &    & radius\,--\,\Rpl & \Rpl  & s$^{-1}$ & s$^{-1}$ &           \\
\hline
\multicolumn{12}{c}{Spectral type: G2 -- \Ms\,=\,1.00\,\Mo\ -- \Rs\,=\,1.00\,\Ro\ -- \Teff\,=\,5777\,K} \\
 5.0 & 2.2 &  2.58 &  500 & 461 &   14.46 & 0.519 & {\bf 34} & 95.0 & 3.2 & 1.0$\times$10$^{32}$ & 6.2$\times$10$^{31}$ &   1.6 \\
 5.0 & 2.4 &  1.99 &  500 & 408 &   14.46 & 0.519 &	 31  & 87.0 & 3.2 & 1.9$\times$10$^{32}$ & 8.0$\times$10$^{31}$ &   2.4 \\
 5.0 & 2.6 &  1.56 &  500 & 364 &   14.46 & 0.519 &	 29  & 80.0 & 3.2 & 3.8$\times$10$^{32}$ & 9.6$\times$10$^{31}$ &   4.0 \\
 5.0 & 3.2 &  0.84 &  500 & 271 &   14.46 & 0.519 &	 24  & 65.3 & 3.5 & 2.4$\times$10$^{33}$ & 2.1$\times$10$^{32}$ &  11.0 \\
 5.0 & 4.4 &  0.32 &  500 & 171 &   14.46 & 0.519 &	 17  & 47.5 & 3.5 & 3.2$\times$10$^{34}$ & 5.6$\times$10$^{32}$ &  57.1 \\
\hline
 5.0 & 1.3 & 12.51 & 1000 & 688 &  231.43 & 0.130 & {\bf 29} & 40.0 & 2.2 & 1.1$\times$10$^{32}$ & 0.9$\times$10$^{32}$ &   1.2 \\
 5.0 & 1.8 &  4.71 & 1000 & 454 &  231.43 & 0.130 &	 21  & 29.0 & 2.1 & 5.9$\times$10$^{32}$ & 2.2$\times$10$^{32}$ &   2.7 \\
 5.0 & 2.3 &  2.26 & 1000 & 331 &  231.43 & 0.130 &	 16  & 22.8 & 2.1 & 2.1$\times$10$^{33}$ & 4.6$\times$10$^{32}$ &   4.6 \\
\hline
10.0 & 4.0 &  0.86 &  500 & 321 &   14.46 & 0.519 & {\bf 38} & 65.8 & 2.4 & 6.3$\times$10$^{31}$ & 1.1$\times$10$^{32}$ &   0.6 \\
10.0 & 5.0 &  0.44 &  500 & 233 &   14.46 & 0.519 &	 30  & 52.6 & 3.0 & 1.7$\times$10$^{33}$ & 3.0$\times$10$^{32}$ &   5.7 \\
10.0 & 7.0 &  0.16 &  500 & 143 &   14.46 & 0.519 &	 22  & 37.6 & 4.5 & 3.8$\times$10$^{34}$ & 1.8$\times$10$^{33}$ &  21.1 \\
\hline
10.0 & 3.2 &  1.68 & 1000 & 336 &  231.43 & 0.130 & {\bf 24} & 20.6 & 1.9 & 4.5$\times$10$^{32}$ & 4.0$\times$10$^{32}$ &   1.1 \\
10.0 & 3.7 &  1.09 & 1000 & 278 &  231.43 & 0.130 &	 20  & 17.8 & 1.9 & 1.2$\times$10$^{33}$ & 7.8$\times$10$^{32}$ &   1.5 \\
10.0 & 4.2 &  0.74 & 1000 & 235 &  231.43 & 0.130 &	 18  & 15.7 & 2.0 & 2.0$\times$10$^{33}$ & 1.1$\times$10$^{33}$ &   1.8 \\
10.0 & 5.0 &  0.44 & 1000 & 186 &  231.43 & 0.130 &	 15  & 13.2 & 2.0 & 5.5$\times$10$^{33}$ & 2.1$\times$10$^{33}$ &   2.6 \\
\hline
\multicolumn{12}{c}{Sp. Type: K2 -- \Ms\,=\,0.76\,\Mo\ -- \Rs\,=\,0.75\,\Ro\ -- \Teff\,=\,5000\,K} \\
 5.0 & 2.6 &  1.56 &  500 & 364 &   25.78 & 0.292 & {\bf 29} & 49.6 & 2.4 & 1.4$\times$10$^{32}$ & 1.1$\times$10$^{32}$ &   1.3 \\
 5.0 & 3.2 &  0.84 &  500 & 271 &   25.78 & 0.292 &	 24  & 40.3 & 2.3 & 3.2$\times$10$^{32}$ & 1.6$\times$10$^{32}$ &   2.0 \\
 5.0 & 3.8 &  0.50 &  500 & 212 &   25.78 & 0.292 &	 20  & 33.9 & 2.3 & 6.8$\times$10$^{32}$ & 2.7$\times$10$^{32}$ &   2.5 \\
 5.0 & 4.4 &  0.32 &  500 & 171 &   25.78 & 0.292 &	 17  & 29.3 & 3.5 & 1.1$\times$10$^{34}$ & 1.0$\times$10$^{33}$ &  11.0 \\
\hline
 5.0 & 1.8 &  4.71 & 1000 & 454 &  412.42 & 0.073 & {\bf 21} & 17.9 & 1.8 & 4.0$\times$10$^{32}$ & 2.9$\times$10$^{32}$ &   1.4 \\
 5.0 & 2.0 &  3.44 & 1000 & 397 &  412.42 & 0.073 &	 19  & 16.1 & 1.9 & 1.1$\times$10$^{33}$ & 4.8$\times$10$^{32}$ &   2.3 \\
 5.0 & 2.3 &  2.26 & 1000 & 331 &  412.42 & 0.073 &	 16  & 14.0 & 1.9 & 1.8$\times$10$^{33}$ & 6.6$\times$10$^{32}$ &   2.7 \\
 5.0 & 2.5 &  1.76 & 1000 & 297 &  412.42 & 0.073 &	 15  & 12.9 & 2.0 & 2.6$\times$10$^{33}$ & 8.5$\times$10$^{32}$ &   3.1 \\
 5.0 & 2.7 &  1.40 & 1000 & 269 &  412.42 & 0.073 &	 14  & 11.9 & 2.0 & 4.9$\times$10$^{33}$ & 1.2$\times$10$^{33}$ &   4.1 \\
 5.0 & 3.5 &  0.64 & 1000 & 191 &  412.42 & 0.073 &	 11  &  9.2 & 2.3 & 2.7$\times$10$^{34}$ & 3.5$\times$10$^{33}$ &   7.7 \\
 5.0 & 4.0 &  0.43 & 1000 & 159 &  412.42 & 0.073 &	  9  &  8.0 & 2.3 & 7.7$\times$10$^{34}$ & 5.1$\times$10$^{33}$ &  15.1 \\
 5.0 & 4.5 &  0.30 & 1000 & 137 &  412.42 & 0.073 &	  8  &  7.2 & 3.2 & 6.0$\times$10$^{35}$ & 1.2$\times$10$^{34}$ &  50.0 \\
\hline
10.0 & 4.0 &  0.86 &  500 & 321 &   25.78 & 0.292 &	 38  & 40.6 & 3.0 & 9.0$\times$10$^{31}$ & 3.0$\times$10$^{32}$ &   0.3 \\
10.0 & 5.0 &  0.44 &  500 & 233 &   25.78 & 0.292 & {\bf 30} & 32.0 & 3.0 & 8.4$\times$10$^{32}$ & 6.0$\times$10$^{32}$ &   1.4 \\
10.0 & 7.0 &  0.16 &  500 & 143 &   25.78 & 0.292 &	 22  & 23.0 & 3.0 & 4.5$\times$10$^{33}$ & 1.5$\times$10$^{33}$ &   3.0 \\
10.0 & 8.0 &  0.11 &  500 & 117 &   25.78 & 0.292 &	 19  & 20.0 & 3.0 & 8.8$\times$10$^{33}$ & 2.2$\times$10$^{33}$ &   4.0 \\
\hline
10.0 & 3.7 &  1.09 & 1000 & 278 &  412.42 & 0.073 &	 20  & 11.0 & 1.6 & 3.1$\times$10$^{32}$ & 1.0$\times$10$^{33}$ &   0.3 \\
10.0 & 4.2 &  0.74 & 1000 & 235 &  412.42 & 0.073 & {\bf 18} &  9.7 & 1.9 & 2.2$\times$10$^{33}$ & 2.0$\times$10$^{33}$ &   1.1 \\
10.0 & 4.7 &  0.53 & 1000 & 202 &  412.42 & 0.073 &	 16  &  8.6 & 1.9 & 4.2$\times$10$^{33}$ & 2.8$\times$10$^{33}$ &   1.5 \\
10.0 & 7.5 &  0.13 & 1000 & 108 &  412.42 & 0.073 &	 10  &  5.4 & 4.0 & 1.9$\times$10$^{36}$ & 7.3$\times$10$^{34}$ &  26.0 \\ 
\hline
\multicolumn{12}{c}{Sp. Type: M2 -- \Ms\,=\,0.40\,\Mo\ -- \Rs\,=\,0.38\,\Ro\ -- \Teff\,=\,3500\,K} \\
 5.0 & 2.6 &  1.56 &  500 & 364 &  107.36 & 0.072 & {\bf 29} & 15.0 & 2.3 & 6.8$\times$10$^{32}$ & 3.8$\times$10$^{32}$ &   1.8 \\
 5.0 & 3.2 &  0.84 &  500 & 271 &  107.36 & 0.072 &	 24  & 12.3 & 2.3 & 1.5$\times$10$^{33}$ & 6.8$\times$10$^{32}$ &   2.2 \\
 5.0 & 3.8 &  0.50 &  500 & 212 &  107.36 & 0.072 &	 20  & 10.3 & 2.3 & 3.7$\times$10$^{33}$ & 1.1$\times$10$^{33}$ &   3.4 \\
\hline
 5.0 & 1.8 &  4.71 & 1000 & 454 & 1717.72 & 0.018 & {\bf 21} &  5.5 & 1.6 & 8.6$\times$10$^{32}$ & 9.4$\times$10$^{32}$ &   0.9 \\
 5.0 & 2.3 &  2.26 & 1000 & 331 & 1717.72 & 0.018 &	 16  &  4.3 & 1.8 & 3.1$\times$10$^{34}$ & 2.5$\times$10$^{33}$ &  12.4 \\
\hline
10.0 & 6.0 &  0.25 &  500 & 179 &  107.36 & 0.072 & {\bf 25} &  8.3 & 2.0 & 1.5$\times$10$^{33}$ & 1.7$\times$10$^{33}$ &   0.9 \\
10.0 & 7.0 &  0.16 &  500 & 143 &  107.36 & 0.072 &	 22  &  7.0 & 2.0 & 8.0$\times$10$^{33}$ & 2.7$\times$10$^{33}$ &   3.0 \\
10.0 & 8.0 &  0.11 &  500 & 117 &  107.36 & 0.072 &	 19  &  6.0 & 2.0 & 1.6$\times$10$^{34}$ & 4.0$\times$10$^{33}$ &   4.0 \\
\hline
10.0 & 3.2 &  1.68 & 1000 & 336 & 1717.72 & 0.018 & {\bf 24} &  3.9 & 1.6 & 4.0$\times$10$^{33}$ & 2.6$\times$10$^{33}$ &   1.5 \\
10.0 & 3.7 &  1.09 & 1000 & 278 & 1717.72 & 0.018 &	 20  &  3.3 & 1.6 & 3.4$\times$10$^{34}$ & 4.2$\times$10$^{33}$ &   8.1 \\
\hline
\end{tabular}
\end{center}
\tablefoot{For all planets we used a pressure at the lower boundary of 100\,mbar.}
\end{table*}

Figure~\ref{fig:table-plot} shows that the \lhy/\len\,$\approx1$ condition is reached for \B\ values between 15 and 35, with a slight dependence on stellar type and \teq. In particular, for the planets orbiting the G- and K-type stars, the \Bc\ values appear to be lower at higher temperature, hence \Bc\ decreases with increasing tidal gravity. This does not seem to be the case for the planets orbiting the M-type star, particularly for the 10\,\Me\ planet.

We discuss here the uncertainties related to the computation of the \lhy/\len\ ratio. Since we do not consider real planets, there are no observational uncertainties connected to the system parameters. The \Reff\ value present in Eq.~(\ref{eq:energyLimited}) is an output of the hydrodynamic code, and it is used to calculate \lhy\ as well. For these reasons, there are no uncertainties on the \Reff\ value. The heating efficiency $\eta$ is therefore the only input parameter for which its uncertainties may affect the \lhy/\len\ ratio. 

Generally, the heating efficiency varies with altitude, and \citet{shematovich2014} concluded that for hot Jupiters the value of $\eta$ in the thermosphere varies between $\approx$10\% and 20\%. Because our model does not self-consistently calculate $\eta$ with height, we assume an average value of 15\% (Sect.~\ref{sec:modelling}). This agrees well with calculations by \citet{owen2012}, who also estimated that $\eta$ values higher than 40\% are unrealistically high. More recently, \citet{salz2016} calculated the average heating efficiency for a set of planets with different masses and radii. They concluded that for planets with $\log(GM_{\rm pl}/R_{\rm pl})$ smaller than 13.11 (the case of the planets considered here), $\eta$ is about 23\%, independent of the planet parameters.

\begin{figure}
\centerline{\includegraphics[width=9.0cm,clip]{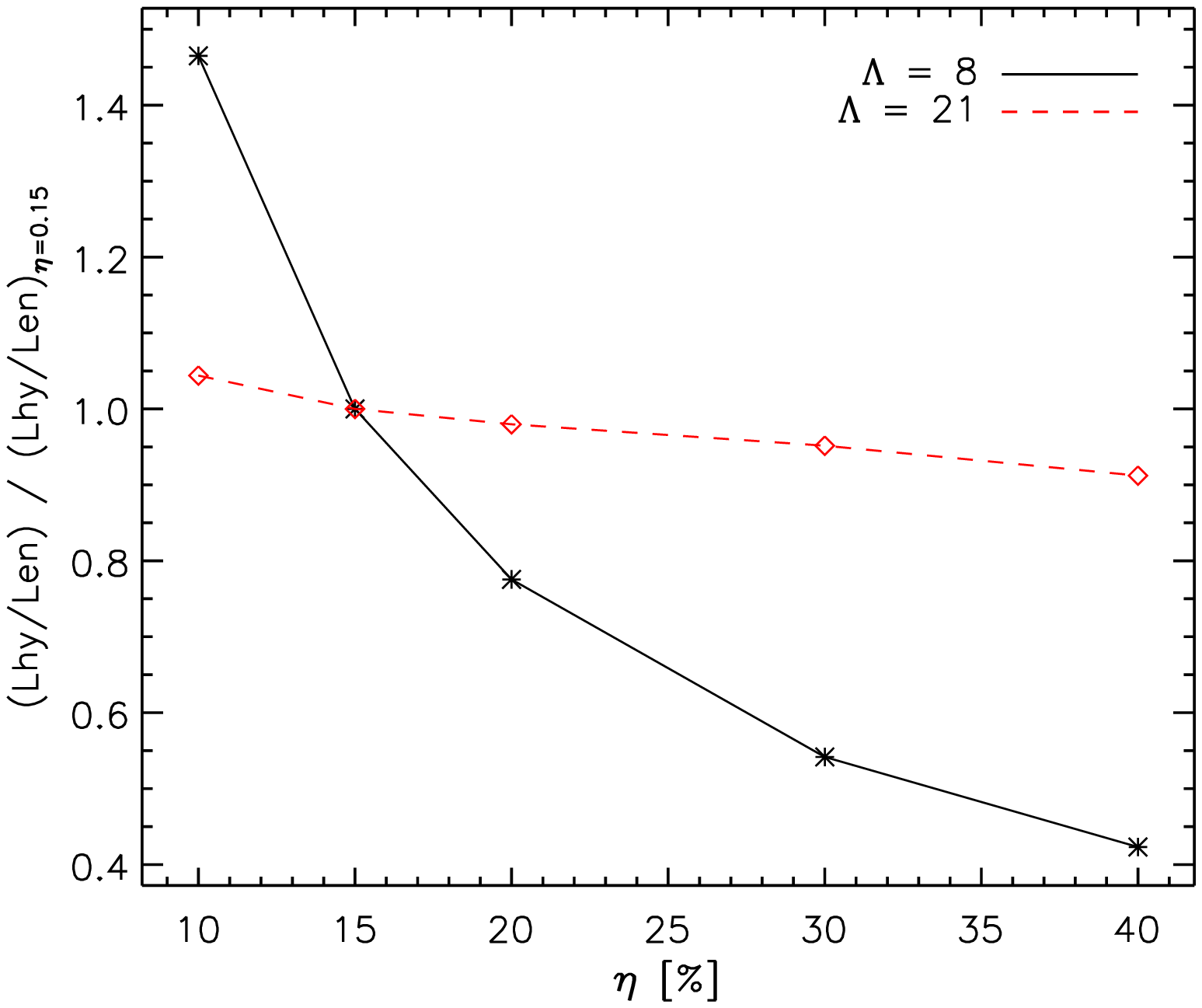}}
\caption{Variation of the \lhy/\len\ ratio, normalised to the value of the \lhy/\len\ ratio obtained with $\eta$\,=\,15\% (adopted for our calculations), as a function of heating efficiency $\eta$ for two planets orbiting the K2 star with two different \B\ values (dashed line: \B\,=\,21, \Mpl\,=\,5\,\Me, \Rpl\,=\,1.8\,\Re; solid line: \B\,=\,8, \Mpl\,=\,5\,\Me, \Rpl\,=\,4.5\,\Re) and \teq\,=\,1000\,K.}
\label{fig:heat_eff}
\end{figure}

As discussed by \citet{lammer2016}, the heating efficiency enters in the calculation of both \lhy\ and \len, although with a slightly different dependence. To quantitatively estimate the effects of the uncertainty on the heating efficiency on the \lhy/\len\ ratio, we ran a set of simulations for two planets orbiting the K2 star with two different \B\ values (\B\,=\,21, \Mpl\,=\,5\,\Me, \Rpl\,=\,1.8\,\Re\ and \B\,=\,8, \Mpl\,=\,5\,\Me, \Rpl\,=\,4.5\,\Re) and \teq\,=\,1000\,K, varying $\eta$ between 10 and 40\%, leaving all other parameters fixed. The results, displayed in Fig.~\ref{fig:heat_eff}, indicate that variations of $\eta$ by a factor of two from the adopted value of 15\% (e.g. between 10 and 30\%) modify the \lhy/\len\ ratio by a factor of about 1.5 in the case of low \B\ and of about 1.05 in the case of high \B. The sensitivity of the \lhy/\len\ ratio on variations of $\eta$ therefore decreases with increasing \B. On the basis of these results, to be conservative, we consider the \lhy/\len\,$\approx1$ condition to be fulfilled when \lhy/\len$\leq2.0$.

\begin{figure}
\centerline{\includegraphics[width=9.0cm,clip]{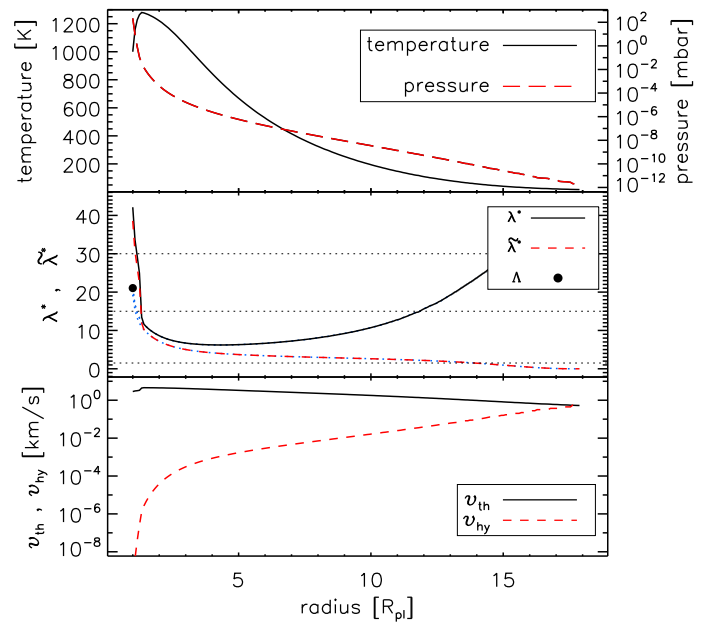}}
\caption{Same as Fig.~\ref{fig:lowbeta}, but for a 5\,\Me\ planet with a radius of 1.8\,\Re.}
\label{fig:highbeta}
\end{figure}

Figure~\ref{fig:highbeta} shows the atmospheric structure of the 5\,\Me\ planet considered in Sect.~\ref{sec:profiles}, but with a radius of 1.8\,\Re\  (i.e. out of the boil-off regime). Close to \Rpl\ the atmosphere is hydrostatic, as indicated by the temperature increase (i.e. no adiabatic cooling), with the high-energy stellar flux providing a considerable amount of heating. The rise in temperature close to the lower boundary in Fig.~\ref{fig:highbeta} is caused by XUV heating, which is the driver of the outflow. In contrast, the monotonic temperature decrease (caused by adiabatic cooling) shown in Fig.~\ref{fig:lowbeta} indicates that XUV heating is not important, implying that the outflow is driven by the high thermal energy of the planet. In our modelling we do not consider cooling from H$_3^+$. However, H$_3^+$ cooling is not relevant in our case, because it does not affect the thermally driven escape rates in the boil-off regime \citep[H$_3^+$ is produced much above the lower boundary of the atmosphere;][]{chadney2016}.

On the basis of detailed evolution modelling of young planets immediately after the disk dispersal, \citet{owen2016} concluded that planets exit the boil-off regime when their radius becomes smaller than 0.1 Bondi radii ($R_{\rm B}$). The Bondi radius is defined as $R_{\rm B}=GM_{\rm pl}/2c_{\rm s}^2$, where $c_{\rm s}$ is the isothermal sound speed. The $R_{\rm pl}/R_{\rm B}=0.1$ condition for the occurrence of boil-off given by \citet{owen2016} is therefore mathematically identical to the \Bc\,$=20$ condition, when an adiabatic gas index $\gamma$ equal to 1 is considered, or in other words, isothermal gas.

We arrived at a result similar to that of \citet{owen2016}, who properly took into account the various heating and cooling sources, which indicates that the assumptions and simplifications we made for our modelling are robust. In particular, it shows the validity of (i) simplifying the processes leading to the planet's thermal balance by setting the temperature of the atmosphere equal to \teq\ at the lower boundary, and (ii) setting the lower boundary at the pressure level where the optical depth is roughly unity, which is where most of the stellar radiation is absorbed\footnote{This is also how \citet{owen2016} set their upper boundary.}. We note that modifications to these two assumptions affect the shape of the atmospheric profiles, but not the escape rates, if \lhy\ is equal to or smaller than \len. For example, Fig.~2 of \citet{lammer2016} shows that by varying the pressure at the lower boundary from 100\,mbar to 1\,bar only affects the \lhy/\len\ ratio in the boil-off regime (when \lhy\,$>$\,\len), while the radius at which the \lhy/\len\,$\approx1$ condition is reached (namely the value of \Bc) is not affected. This implies that, within our scheme, the \Bc\ values are independent of the two assumptions described above. It should also be noted that our results apply to any planet, independent of the internal structure, for which the 100\,mbar pressure level lies above the solid core, if any is present.
\section{Constraints on \Mpl\ and \Rpl}\label{sec:constraints}
To explore whether the knowledge of the value of \Bc, or equivalently of the $R_{\rm pl}/R_{\rm B}=0.1$ condition, can help to constrain the parameters of old planets, it is necessary to consider the atmospheric evolution of planets in the boil-off regime. To roughly estimate how much time the modelled planets need to evolve out of the boil-off regime, we follow the same procedure as adopted by \citet{lammer2016} to study the case of CoRoT-24b. 

\begin{figure}
\centerline{\includegraphics[width=9.0cm,clip]{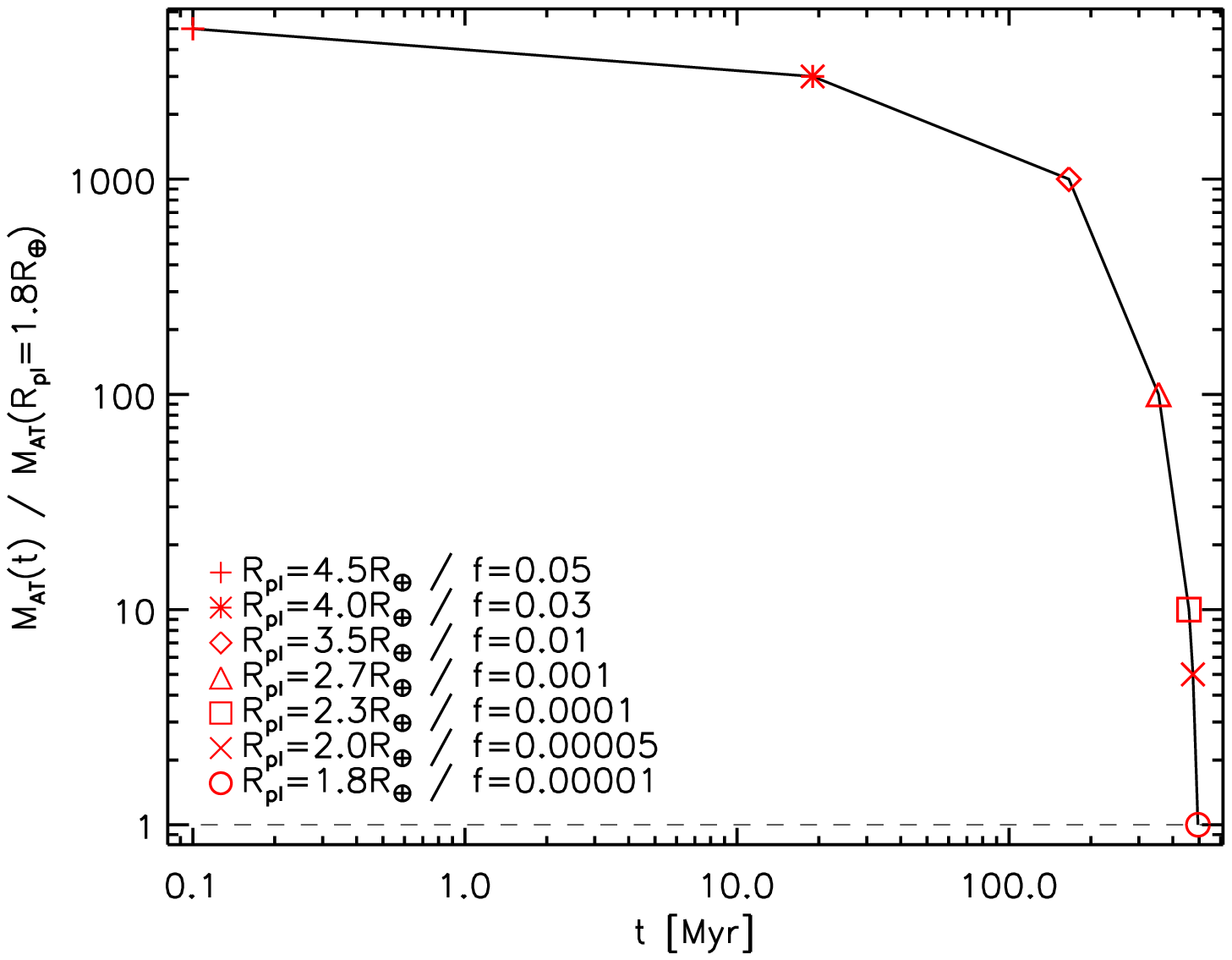}}
\caption{Atmospheric mass $M_{\rm AT}$ evolution normalised to the atmospheric mass corresponding to \Rpl\,=1.8\,\Re\  (where \lhy/\len\,$\approx1$) estimated from the \lhy\ escape rates obtained for the \Mpl\,=\,5\,\Me\ planet with \teq\,=\,1000\,K orbiting the K-type star. The dashed line indicates $M_{\rm AT}$\,=\,$M_{\rm AT}$(1.8\,\Re). The initial time is arbitrarily set at 0.1\,Myr. The legend lists the atmospheric mass fraction corresponding to each radius.}
\label{fig:evolution}
\end{figure}

As an example, we take the simulations we carried out for the \Mpl\,=\,5\,\Me\ planet with \teq\,=\,1000\,K orbiting the K-type star. We assumed a core mass of 5\,\Me\ and used formation and structure models by \citet[][see their Fig.~4]{rogers2011} to estimate for each modelled radius the atmospheric mass fraction $f$. We then used the \lhy\ values to roughly estimate the evolution of the atmospheric mass over time. Figure~\ref{fig:evolution} shows that the atmospheric mass for a radius above 1.8\,\Re\  (where \lhy/\len\,$\approx1$) would be lost within $\approx$500\,Myr. This is therefore the timescale needed for this planet to evolve out of the boil-off regime.

Table~\ref{tab:ages} lists the timescales for each modelled planet from Table~\ref{tab:runs}. We find the shorter time scales for the less massive and hotter planets. In particular, for planets with \Mpl\,=\,5\,\Me\ and \teq\,=\,1000\,K, the timescale to evolve out of boil-off is shorter than 500\,Myr. The same also occurs for the hot (i.e. \teq\,=\,1000\,K) 10\,\Me\ planet orbiting the M-type star, likely because of the effect of the smaller Roche-lobe radius compared to the case of the same planet orbiting the G- and K-type stars. In general, we therefore find that hot (i.e. \teq\,$\gtrapprox$\,1000\,K) low-mass (\Mpl\,$\lessapprox$\,5\,\Me) planets with hydrogen-dominated atmospheres, unless very young, should not have \B\,$<$\,\Bc. Because of their small Roche lobe, this conclusion also extends to hot (i.e. \teq\,$\gtrapprox$\,1000\,K) higher mass (\Mpl\,$\lessapprox$\,10\,\Me) planets if they are orbiting M-dwarfs.
\begin{table}
\caption{Approximate time scales (in Myr) needed for the modelled planets to evolve out of the boil-off regime, following the analysis described in Sect.~\ref{sec:constraints}. Timescales larger than 1\,Gyr have been rounded to the nearest 100\,Myr. The fourth and fifth columns indicate the initial and final \B\ values of the planets used to calculate the timescales.}
\label{tab:ages}
\begin{center}
\begin{tabular}{cccccr}
\hline
Star & \Mpl & \teq & \B$_{\rm i}$ & \B$_{\rm f}$ & Time scale \\
     & \Me  &  K   &              &              & Myr  \\
\hline
G2 &  5 &  500 & 17 & 34 &  14,600 \\
G2 &  5 & 1000 & 16 & 29 &      48 \\
G2 & 10 &  500 & 22 & 38 &  72,500 \\
G2 & 10 & 1000 & 15 & 24 &  22,700 \\
\hline
K2 &  5 &  500 & 17 & 29 &  14,600 \\
K2 &  5 & 1000 &  8 & 21 &     495 \\
K2 & 10 &  500 & 19 & 30 & 102,200 \\
K2 & 10 & 1000 & 10 & 18 &   8,200 \\
\hline
M2 &  5 &  500 & 20 & 29 &  28,000 \\
M2 &  5 & 1000 & 16 & 21 &       3 \\
M2 & 10 &  500 & 19 & 25 &  24,800 \\
M2 & 10 & 1000 & 20 & 24 &     167 \\
\hline
\end{tabular}
\end{center}
\end{table}

From the above considerations, it follows that for hot low-mass planets with hydrogen-dominated atmospheres with observed values leading to \B\,$<$\,15--35 there must be problems with the estimation/interpretation of the measured mass (i.e. too low), or radius (i.e. too large), or both. Large transit radii may be caused by the presence of aerosols lying far above \Rpl\ or by an incorrect estimation of the stellar radius. We note, however, that the atmosphere of planets with a large enough atmospheric mass may stably lie in the boil-off regime, as described above.
\begin{figure}
\centerline{\includegraphics[width=9.0cm,clip]{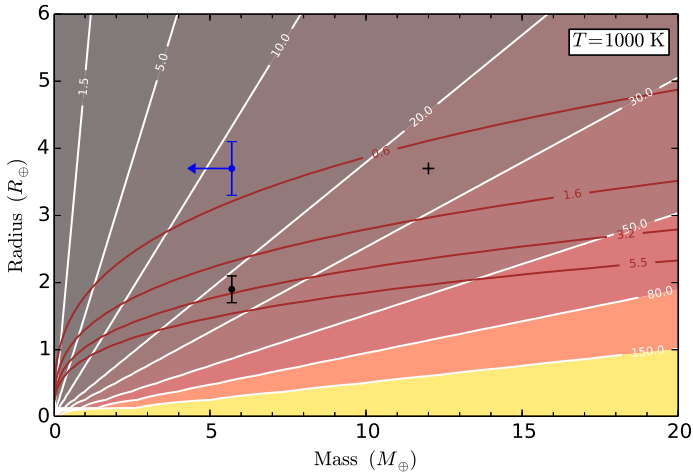}}
\caption{Colour-scaled value of \B\ as a function of planetary mass and radius for \teq\,=\,1000\,K. The white straight lines indicate equal \B\ values given in the plot. The red solid lines indicate lines of equal average densities of 0.6, 1.6, 3.2, and 5.5\,g\,cm$^{-3}$. The symbols correspond to the observed (blue bar and arrow) and possible mass-radius combination (black points) for CoRoT-24b.}
\label{fig:beta}
\end{figure}

The presence of aerosols may indeed lead to a misinterpretation of the observed transit radius. \citet{clouds2015}, for example, calculated from first principles the formation of aerosols in the atmosphere of the hot Jupiter HD\,189733\,b (\teq\,$\approx$1000\,K, similar to that of the hottest planets considered in this work), obtaining that clouds start forming in the 10--100\,$\mu$bar pressure range. For the planet considered in Fig.~\ref{fig:highbeta} (\Mpl\,=\,5\,\Me; \Rpl\,=\,1.8\,\Re; \teq\,=\,1000\,K), this pressure level corresponds to about 1.2--1.4\,\Rpl, that is, a radius of 2.2--2.5\,\Re\ or 5.3--9.3 pressure scale heights above \Rpl. The presence of high-altitude clouds/hazes in the atmosphere of such a planet would therefore lead to an overestimation of \Rpl\ measured through broad-band optical transit observations of about 20--40\%. \citet{clouds2015} investigated a hot Jupiter, which has physical characteristics different from those of the planets considered here, but this is what is currently available, showing that similar cloud formation calculations, tuned for lower-mass planets, are clearly needed for a more appropriate interpretation of the results.

For hot low-mass planets it is therefore possible to use the \B\,$\geq$\,\Bc\ condition to constrain the minimum mass, given a certain radius, or maximum radius, given a certain mass. The only assumption is the presence of a hydrogen-dominated atmosphere, which is likely for low-density planets, and an old age (i.e. $>$\,1\,Gyr). Most of the extremely low-density planets discovered by Kepler fall into this regime. 

Figure~\ref{fig:beta} shows the \B\ value as a function of planetary mass and radius (at the 100\,mbar level) for \teq\,=\,1000\,K. We use the sub-Neptune CoRoT-24b as an example of the constraining power of this plot. CoRoT-24b has a mass lower than 5.7\,\Me, a transit radius of 3.7$\pm$0.4\,\Re, and an equilibrium temperature of 1070\,K \citep[blue bar and arrow in Fig.~\ref{fig:beta};][]{alonso2014}. CoRoT-24b therefore has a \B\ value lower than 10.9, well below \Bc. For a value of \Bc\ of 25 and when we assume that \Mpl\ is equal to 5.7\,\Me\ \citep[][excluded masses smaller than $\approx$5\,\Me]{lammer2016}, Fig.~\ref{fig:beta} (bottom black point) indicates that the 100\,mbar pressure level, and hence where the transit radius would be if the planet were possessed of a clear atmosphere, lies around 2\,\Re\  ($\approx$1.7\,\Re\ less than the transit radius), in agreement with the detailed analysis of \citet{lammer2016}. When we instead assume a clear atmosphere, hence \Rt\,=\,$R_{\rm 100\,mbar}$, \Mpl\ should be $\gtrapprox12$\,\Me\  (right black cross in Fig.~\ref{fig:beta}), although this is unlikely given the non-detection of the planet in the radial-velocity measurements.

The atmospheric pressure profile of CoRoT-24b shown by \citet{lammer2016} indicates that if we assume that the 100\,mbar level lies at 2\,\Re, then the transit radius is at a pressure of 1--10\,$\mu$bar, which is about 10 times smaller than the lowest pressure at which \citet{clouds2015} predicts cloud formation. For this particular planet, the most likely scenario is therefore a combined effect of the presence of aerosols and of a slight mass underestimation.    

Table~\ref{tab:ages} shows that for most of the more massive planets (\Mpl\,$\gtrapprox$10\,\Me) and all the cooler (\teq\,$\lessapprox$\,500\,K) ones, the timescale for the atmosphere to evolve out of the boil-off regime is longer than 10\,Gyr and in some cases even longer than the main-sequence life time of the host stars. This clearly shows that although the atmosphere of these planets may be in boil-off, the escape rates are not high enough to significantly affect the atmosphere in a short time, in agreement with the results of \citet{ginzburg2015}.

From the results of Table~\ref{tab:ages}, it follows that in the 5--10\,\Me\ planetary mass and 500--1000\,K equilibrium temperature range with increasing temperature and/or decreasing mass the escape rates start affecting the long-term evolution of the atmosphere. This transition region depends not only on the planetary parameters, but also on the stellar properties and orbital separation, which affect the escape rates through the XUV flux and size of the Roche lobe. We will explore this transition region in detail in a forthcoming work.
\section{Conclusions}\label{sec:conclusions}
We generalised the expression of the Jeans escape parameter to account for hydrodynamic and Roche-lobe effects, which is important for close-in exoplanets. We use a planetary upper atmosphere hydrodynamic code to derive the atmospheric temperature, pressure, and velocity structure of sub-Neptunes with various masses and radii and draw the profiles of the Jeans escape parameter as a function of height. We used our simulations and the generalised Jeans escape parameter to describe the boil-off regime \citep{owen2016}, which is characterised by very high escape rates driven by the planet's high thermal energy and low gravity.

We introduce the restricted Jeans escape parameter (\B) as the value of the Jeans escape parameter calculated at the observed planetary radius and mass for the planet's equilibrium temperature, and considering atomic hydrogen. We used the \lhy/\len\,$\leq1$ empirical condition, where \len\ is derived analytically from the energy-limited formula, to estimate \Bc, the critical value of \B\ below which efficient boil-off occurs. We ran simulations with varying planetary mass, stellar mass, and equilibrium temperature, concluding that \Bc\ lies between 15 and 35, depending on the system parameters. This result, mostly based on aeronomical considerations, is in agreement with that obtained by \citet{owen2016}, namely $R_{\rm pl}/R_{\rm B}>0.1$.

From the analysis of our simulations, we find that the atmosphere of hot (i.e. \teq\,$\gtrapprox$\,1000\,K) low-mass (\Mpl\,$\lessapprox$\,5\,\Me) planets with \B\,$<$\,\Bc\ would be unstable against evaporation because they lie in an efficient boil-off regime that would shrink their radius within  a few hundreds of Myr. We find the same result also for hot (i.e. \teq\,$\gtrapprox$\,1000\,K) higher mass (\Mpl\,$\lessapprox$\,10\,\Me) planets with \B\,$<$\,\Bc, when they orbit M-dwarfs. We conclude that for old hydrogen-dominated planets in this range of parameters, \B\ should be $\geq$\,\Bc, which therefore provides a strong constraint on the planetary minimum mass/maximum radius.

This information can be used to predict the presence of high-altitude aerosols on a certain planet without the need to obtain transmission spectra, or inform on the reliability of planetary masses. Our results could also be used to indicate the possible presence of contaminants in the images used to derive the transit light curves, which would lead to the measurement of a planetary radius larger than what is in reality \citep{dalba2016}. Our results are relevant because of the various present and future ground- and space-based planet-finding facilities (e.g. K2, NGTS, CHEOPS, TESS, PLATO), which will detect sub-Neptunes orbiting bright stars, hence amenable to atmospheric characterisation. Our results will help prioritisation processes: for instance, hot low-density, low-mass planets, with masses measured through radial velocity, are good targets for transmission spectroscopy, but their large radii may be caused by high-altitude clouds, which would therefore obscure the atmospheric atomic and molecular features. An application of our results to the transiting sub-Neptune planets known to date is presented by \citet{cubillos2016}.

The simulations presented in this work, only sparsely cover the typical parameter space of the discovered systems hosting sub-Neptunes, also in terms of high-energy stellar flux. In the future, we will extend our work to a larger parameter space and aiming at its more homogeneous coverage. In particular, we will better identify the dependence of the \Bc\ value on the planetary (e.g. mass, radius, and temperature/pressure at the lower boundary) and stellar (e.g. mass and high-energy flux) parameters.
\begin{acknowledgements}
We acknowledge the Austrian Forschungsf\"orderungsgesellschaft FFG projects ``RASEN'' P847963 and ``TAPAS4CHEOPS'' P853993, the Austrian Science Fund (FWF) NFN project S11607-N16, and the FWF project P27256-N27. NVE acknowledges support by the RFBR grant No. 15-05-00879-a and 16-52-14006 ANF\_a. We thank the anonymous referee for the comments that led to a considerable improvement of the manuscript.
\end{acknowledgements}
\end{document}